\documentclass[12pt, a4]{article}
\usepackage{amsfonts}
\usepackage{graphicx}
\usepackage{amsmath}
\usepackage{amssymb}
\usepackage{subcaption}

\begin{document}

\title{\textbf{Complete Integrability of a New Class of Hamiltonian
Hydrodynamic Type Systems}}
\author{Zakhar V. Makridin$^1$, Maxim V. Pavlov$^2$ \\
[-3pt] {\small $^1$ Lavrentyev Institute of Hydrodynamics,}\\ 
[-3pt] {\small Lavrentyev street 15, 630090, Novosibirsk, Russia}\\
[-3pt] {\small $^2$ Lebedev Physical Institute of Russian Academy of Sciences,}\\
[-3pt] {\small	Leninskij Prospekt 53, 119991 Moscow, Russia }}
\date{\today }
\maketitle

\begin{abstract}
In this paper we consider a new class of Hamiltonian hydrodynamic type
systems, whose conservation laws are polynomial with respect to one of field
variables.
\end{abstract}

\tableofcontents

\newpage

\section{Introduction}

The theory of Hamiltonian hydrodynamic type systems%
\begin{equation*}
u_{t}^{k}=\eta ^{km}\left( \frac{\partial h}{\partial u^{m}}\right)
_{x},\quad k,m=1,2,...,N
\end{equation*}%
was initiated from the seminal paper \cite{dn}, written by B.A. Dubrovin and
S.P. Novikov in 1983. Here the matrix $\eta ^{ik}$ is constant, while a
Hamiltonian density $h(u)$ is an arbitrary function of $N$ field variables $%
u^{k}$ only. If $N=2$, this is a \textquotedblleft
classical\textquotedblright\ case, because such systems are
\textquotedblleft linearisable\textquotedblright\ by a hodograph method.
\textquotedblleft Linearisation\textquotedblright\ means transformation of a
quasilinear system to a linear system with variable coefficients. In this
paper we are going to consider the special Hamiltonian case%
\begin{equation}
u_{t}=\left( \frac{\partial h}{\partial v}\right) _{x},\quad v_{t}=\left( 
\frac{\partial h}{\partial u}\right) _{x},  \label{ham}
\end{equation}%
where a Hamiltonian density $h$ depends on one of the field variables $v$ 
\textit{quadratically}, i.e.%
\begin{equation}
h(u,v)=A(u)v^{2}+B(u)v+C(u).  \label{nonsep}
\end{equation}%
Here $A(u),B(u)$ and $C(u)$ are arbitrary analytic functions. We show that
such Hamiltonian systems possess infinitely many particular solutions, which
can be found by integration of \textit{linear ordinary differential
equations of second order} only.

The class of Hamiltonian densities \eqref{nonsep} being considered is
meaningful since it consist of some physically relevant examples will be
discussed in subsequent sections:

\begin{itemize}
\item The shallow water equations with densities 
\begin{equation*}
h_{ESWE}=\frac{1}{2}uv^2+C(u),\quad h_{LSWE}=\frac{1}{2}v^2+C(u)
\end{equation*}
corresponding to Eulerian and Lagrangian coordinates respectively;

\item The two-layer non-Boussinesq shallow water equations \cite{ovsyannikov}%
, \cite{camassa} with density 
\begin{equation*}
h_{TSWE}=-\frac{1}{4}\frac{(1-u^2)}{(1-r u)}v^2-\frac{1}{4}u^2,
\end{equation*}
where $r=const$ is the Boussinesq parameter;

\item The system, describing motions of liquid layer on a rotating cylinder 
\cite{MZH}: 
\begin{equation*}
h_{LLRC}=-\frac{1}{2}v^2\ln{u}+\frac{1}{2}v\,\Omega u\ln{u}-\frac{1}{8}%
\Omega^2u^2\ln{u}+\frac{1}{4}\Omega^2u^2
\end{equation*}
with $\Omega$ being constant vorticity.
\end{itemize}

Since any two-component hydrodynamic type system is lineariable by the
hodograph method, Hamiltonian system (\ref{ham}) possesses infinitely many
conservation laws $(f(u,v))_{t}=(g(u,v))_{x}$, where conservation law
densities $f(u,v)$ satisfy a second order linear equation%
\begin{equation}
h_{uu}f_{vv}=h_{vv}f_{uu},  \label{trick}
\end{equation}%
whose general solution is parameterised by two arbitrary functions of a
single variable. However, this general solution is impossible to find in
explicit form for an arbitrary function $h(u,v)$. By this reason, a
possibility to construct infinitely many particular solutions also is very
important. The first nontrivial case was found by Ya. Nutku and P. Olver
(see \cite{on}). They selected the so called \textit{separable} class%
\begin{equation*}
p(v)f_{vv}=q(u)f_{uu},
\end{equation*}%
which is connected with two types of Hamiltonian densities $h=A(u)B(v)$ and $%
h=A(u)+B(v)$. Later Ya. Nutku and P. Olver formulated the question
\textquotedblleft \textit{how many of results presented for the separable
class can be generalised to the \textbf{nonseparable} classes}%
?\textquotedblright

In this paper we present a first example belonging to non-separable classes,
selected by the choice of Hamiltonian density (\ref{nonsep}). Then linear
equation (\ref{trick}) reduces to the form%
\begin{equation}
(A^{\prime \prime }(u)v^{2}+B^{\prime \prime }(u)v+C^{\prime \prime
}(u))f_{vv}=2A(u)f_{uu}.  \label{omi}
\end{equation}%
\textbf{Our claim is} that this \textit{linear PDE} has infinitely many
particular polynomial solutions with respect to an independent variable $v$,
selected by \textit{infinitely many} corresponding \textit{second order ODEs}%
. This relies on the following simple observation. Any linear equation (\ref%
{trick}) has two particular polynomial solutions of first degree $f=u$ and $%
f=v$. Now, consider ansatz 
\begin{equation*}
f=G_{2,2}(u)v^2+G_{2,1}(u)v+G_{2,0}(u),
\end{equation*}
then \eqref{omi} decouples on three ODEs for functions $G_{2,i}$: 
\begin{equation}  \label{g22}
AG_{2,2}^{\prime \prime }=A^{\prime \prime }G_{2,2},\quad AG_{2,1}^{\prime
\prime }=B^{\prime \prime }G_{2,2},\quad AG_{2,0}^{\prime \prime }=C^{\prime
\prime }G_{2,2},
\end{equation}
which can be integrated immediately, giving two particular solutions, and
one of them is $f=h$. Then let us take cubic polynomial: 
\begin{equation*}
f=G_{3,3}(u)v^{3}+G_{3,2}(u)v^{2}+G_{3,1}(u)v+G_{3,0}(u).
\end{equation*}%
The corresponding ODEs are 
\begin{equation}  \label{g33}
AG_{3,3}^{\prime \prime }=3 A^{\prime \prime }G_{3,3},\quad AG_{3,2}^{\prime
\prime }=A^{\prime \prime }G_{3,2}+3B^{\prime \prime }G_{3,3},
\end{equation}
\begin{equation*}
AG_{3,1}^{\prime \prime }=B^{\prime \prime }G_{3,2}+3C^{\prime \prime
}G_{3,3},\quad AG_{3,0}^{\prime \prime }=C^{\prime \prime }G_{3,2}.
\end{equation*}
In the same way, substituting fourth order conservation law densities%
\begin{equation*}
f=G_{4,4}(u)v^{4}+G_{4,3}(u)v^{3}+G_{4,2}(u)v^{2}+G_{4,1}(u)v+G_{4,0}(u)
\end{equation*}
into (\ref{omi}) lead to the system 
\begin{equation}  \label{g44}
AG_{4,4}^{\prime \prime }=6A^{\prime \prime }G_{4,4},\quad AG_{4,3}^{\prime
\prime }=3A^{\prime \prime }G_{4,3}+6B^{\prime \prime }G_{4,4},
\end{equation}
\begin{equation*}
AG_{4,2}^{\prime \prime }=A^{\prime \prime }G_{4,2}+3B^{\prime \prime
}G_{4,3}+6C^{\prime \prime }G_{4,4},\quad AG_{4,1}^{\prime \prime
}=B^{\prime \prime }G_{4,2}+3C^{\prime \prime }G_{4,3}, \quad
AG_{4,0}^{\prime \prime }=C^{\prime \prime }G_{4,2}.
\end{equation*}
So, one can see that the first ODE in \eqref{g22}, \eqref{g33} and %
\eqref{g44} is the same up to the factor of $A^{\prime \prime }$. Also, it
is seen, that the second ODE in \eqref{g33} and \eqref{g44} has the same
homogeneous part as the first one in \eqref{g22} and \eqref{g33}
respectively. This means, that once the first equation in \eqref{g22}, %
\eqref{g33} and \eqref{g44} is solved, one can integrate all remaining ODEs
in quadratures.

Further investigation of other higher polynomial conservation law densities
(with respect to an independent variable $v$) shows that: if the first
linear ODE (here $k$ is a natural number determining a degree of a
corresponding polynomial conservation law density with respect to an
independent variable $v$)%
\begin{equation*}
AG_{k,k}^{\prime \prime }=\frac{1}{2}k(k-1)A^{\prime \prime }G_{k,k}
\end{equation*}%
can be solved, then other coefficients $G_{k,i}$ for $i<k$ can be found in
quadratures. Below we describe our construction in details.

\section{Integrability and Complete Integrability}

Quasilinear system (\ref{ham}) under the hodograph transform $%
(u,v)\leftrightarrow (x,t)$ becomes a linear system%
\begin{equation}
x_{v}=t_{u}h_{vv}-t_{v}h_{uv},\quad x_{u}=t_{v}h_{uu}-t_{u}h_{uv},
\label{eks}
\end{equation}%
with variable coefficients $h_{uu},h_{uv},h_{vv}$. Eliminating $x$, one can
obtain a linear equation of second order%
\begin{equation}
2t_{u}h_{uvv}-2t_{v}h_{uuv}+t_{uu}h_{vv}-t_{vv}h_{uu}=0,  \label{lina}
\end{equation}%
with variable coefficients $h_{uu},h_{vv},h_{uuv},h_{uvv}$.

The Tsarev generalised hodograph method (see \cite{tsar}) is based on a
concept of commuting flows. This means, that Hamiltonian system (\ref{ham})
possesses infinitely many commuting flows%
\begin{equation*}
u_y=\left(\frac{\partial f}{\partial v}\right)_x,\quad v_y=\left(\frac{%
\partial f}{\partial u}\right)_x,
\end{equation*}%
where $y$ is the so-called group parameter and a Hamiltonian density $f(u,v)$
cannot be arbitrary. Indeed, the compatibility conditions $%
(u_{y})_{t}=(u_{t})_{y}$, $(v_{y})_{t}=(v_{t})_{y}$ lead to a single second
order linear equation%
\begin{equation}
h_{uu}f_{vv}=h_{vv}f_{uu}.  \label{linb}
\end{equation}%
In this case, solutions of both linear equations (\ref{lina}), (\ref{linb})
are connected by equivalent substitutions%
\begin{equation}  \label{t}
t=\frac{f_{vv}}{h_{vv}}=\frac{f_{uu}}{h_{uu}}.
\end{equation}%
Then the dependence $x(u,v)$ can be found by quadratures from (\ref{eks}):%
\begin{equation}  \label{x}
x=f_{uv}-t h_{uv}.
\end{equation}%
So, in this paper we consider more simple equation (\ref{linb}) than (\ref%
{lina}). As it was mentioned in previous section we have a second order
linear hyperbolic equation, whose general solution depends on two arbitrary
functions of a single variable. Now, we define our concept of solvability:

\textit{Equation} (\ref{linb}) \textit{is called solvable, if it possesses
infinitely many particular solutions, where each of them can be found from
corresponding linear \textbf{ordinary} differential equation of second order}%
.

Assume that we are able to construct a general solution, parameterised by
two arbitrary functions of a single variable. Then we can expand these two
functions in a Taylor series. This means, we are able to construct two
infinite series of particular solutions. Viceversa, if we are able to
construct two infinite series of particular solutions, we expect that they
can be summated to a general solution, parameterised by two arbitrary
functions of a single variable. So, here we give our definition of
completeness:

\textit{Equation} (\ref{linb}) \textit{is called completely solvable, if it
possesses two infinite series of particular solutions, where each of them
can be found from corresponding linear \textbf{ordinary} differential
equation of second order}.

According to our definition, equation \eqref{linb} is completely solvable,
when function $h$ is of the form 
\begin{equation*}
h(u,v)=A(u)v^{2}+B(u)v+C(u)
\end{equation*}%
with arbitrary analytic functions $A$, $B$ and $C$. To prove this statement
one should look for solution of the linear equation \eqref{linb} as a
polynomial with arbitrary degree $k$: 
\begin{equation}
f_{k}(u,v)=G_{k,k}(u)v^{k}+\ldots +G_{k,1}(u)v+G_{k,0}(u).  \label{fg}
\end{equation}%
Substituting this formula in \eqref{linb} we obtain the following system of
inhomogeneous second order ordinary differential equation for coefficients $%
G_{k,i}(u)$: 
\begin{align}
& 2AG_{k,k}^{\prime \prime }-k(k-1)A^{\prime \prime }G_{k,k}=0,  \notag
\label{ode} \\
& 2AG_{k,k-1}^{\prime \prime }-(k-1)(k-2)A^{\prime \prime
}G_{k,k-1}=k(k-1)B^{\prime \prime }G_{k,k},  \notag \\
& 2AG_{k,k-2}^{\prime \prime }-(k-2)(k-3)A^{\prime \prime
}G_{k,k-2}=(k-1)(k-2)B^{\prime \prime }G_{k,k-1}  \notag \\
& \qquad \qquad \qquad \qquad \qquad \qquad \qquad \qquad \qquad
+k(k-1)C^{\prime \prime }G_{k,k}, \\
& \,  \notag \\
& 2AG_{k,k-s}^{\prime \prime }-(k-s)(k-s-1)A^{\prime \prime
}G_{k,k-s}=(k-s)(k-s+1)B^{\prime \prime }G_{k,k-s+1}  \notag \\
& \qquad \qquad \qquad \qquad \qquad \qquad \qquad \qquad
+(k-s+1)(k-s+2)C^{\prime \prime }G_{k,k-s+2},  \notag
\end{align}%
where $s=3,\ldots ,k.$ Consider the first equation of the system \eqref{ode}%
: 
\begin{equation}
2AG_{k,k}^{\prime \prime }-k(k-1)A^{\prime \prime }G_{k,k}=0.  \label{gk}
\end{equation}%
Suppose that analytic function $A(u)$ is such that the equality holds: 
\begin{equation*}
\frac{A^{\prime \prime }(u)}{A(u)}=\frac{a_{-2}}{(u-u_{0})^{2}}+\frac{a_{-1}%
}{(u-u_{0})}+a_{0}+a_{1}(u-u_{0})+\ldots
\end{equation*}%
for some vicinity of point $u_{0}$. It means that point $u_{0}$ is weakly
singular (or regular when $a_{-2}=a_{-1}=0$) for equation \eqref{gk}.
According to the general analytical theory of ODEs \cite{liz} there exist a
particular solution which can be represented as a convergent series 
\begin{equation*}
G_{k,k}^{1}(u)=(u-u_{0})^{\gamma }\sum\limits_{m=1}^{\infty
}g_{m}(u-u_{0})^{m},
\end{equation*}%
where exponent $\gamma $ are defined as a bigger root of the characteristic
equation 
\begin{equation*}
a_{-2}-\gamma (\gamma -1)=0.
\end{equation*}%
The second linearly independent solution $G_{k,k}^{2}(u)$ is also
represented as a convergent series and its form depends on properties of
parameter $\gamma $ (see details in \cite{liz}). Thus, having obtained the
solution of equation \eqref{gk} in the form 
\begin{equation*}
G_{k,k}(u)=c_{1}G_{k,k}^{1}(u)+c_{2}G_{k,k}^{2}(u),
\end{equation*}%
one can easily construct the solution of whole system \eqref{ode}.%
\begin{equation}
AG_{k,0}^{\prime \prime }=C^{\prime \prime }G_{k,2},\text{ \ }%
AG_{k,1}^{\prime \prime }=B^{\prime \prime }G_{k,2}+3C^{\prime \prime
}G_{k,3}.  \notag
\end{equation}

\section{Two-Layer Non-Boussinesq Shallow Water System}

Consider the following system 
\begin{align}  \label{xs}
& \xi_t=-\left\{\frac{(h^2-\xi^2)\sigma}{(\rho_2+\rho_1)h-(\rho_2-\rho_1)\xi}%
\right\}_x \\[3.5mm]
&\sigma_t=-\frac{1}{2}\left\{\frac{\rho_2(h-\xi)^2-\rho_1(h+\xi)^2}{%
((\rho_2+\rho_1)h-(\rho_2-\rho_1)\xi)^2}\sigma^2+g(\rho_2-\rho_1)\xi\right%
\}_x,  \notag
\end{align}
arising in describing of long small-amplitude nonlinear waves in
incompressible two-layered flows in infinite channel \cite{ovsyannikov},\cite%
{camassa}. Here $\rho_{1,2}$ is the density in upper and lower layer
correspondingly, $h$ is the total height of the channel, $\xi$ is the
relative thickness (i.e. the difference between layer thicknesses) and $%
\sigma$ is the momentum shear. In dimensionless variables (see details in 
\cite{camassa}) the system \eqref{xs} takes the form \eqref{ham} with
Hamiltonian density 
\begin{align}  \label{htwo}
h(u,v;r)=-\frac{1}{4}\frac{(1-u^2)}{(1-r u)}v^2-\frac{1}{4}u^2,
\end{align}
where $r=(\rho_2-\rho_1)/(\rho_2+\rho_1)$ is the Boussinesq parameter.

There are two limit cases:

\begin{itemize}
\item \textbf{the first case}, when $r\to 1$, corresponds $\rho_1\to 0$,
i.e. zero density in the upper layer. Therefore the interface can be
considered as a free surface and our system transform to the shallow water
model up to point transformation. Indeed, taking $r\to 1$ the Hamiltonian
density \eqref{htwo} becomes 
\begin{equation*}
h(u,v;1)=-\frac{1}{4}(1+u)v^2-\frac{1}{4}u^2,
\end{equation*}
which upon the substitution $u=2U-1$, $v=V$ gives the shallow water system.
The corresponding equation \eqref{linb} takes the following form 
\begin{equation*}
f_{VV}-Uf_{UU}=0,
\end{equation*}
which has infinitely many polynomial solutions (see \cite{whitham}).

\item \textbf{the second case}, $r\to 0$, related to the so-called
Boussinesq approximation, means that densities in both layers are nearly
coincide. In terms of parameter $r$ the relation $\rho_1\approx\rho_2$
corresponds to $r\to 0$. Surprisingly, but this case is also related to the
shallow water system of equations. As noted in \cite{ovsyannikov} (see also 
\cite{gavsukhmak}) the system \eqref{ham} with density 
\begin{equation*}
h(u,v;0)=-\frac{1}{4}(1-u^2)v^2-\frac{1}{4}u^2,
\end{equation*}
coincide, under the map 
\begin{equation*}
\eta=(1-u^2)(1-v^2), \quad \zeta=2uv,
\end{equation*}
with shallow water equations, written for field variables $\eta$ and $\zeta$%
. Thus, the solutions of two-layer non-Boussinesq shallow water system are,
in some sense, deformations of those, constructed for ordinary shallow water
equations \cite{camassa}.
\end{itemize}

Turning back to the Hamiltonian density \eqref{htwo}, we have explicit
expressions for functions $A$, $B$ and $C$: 
\begin{equation*}
A(u)=-\frac{1}{4}\frac{(1-u^2)}{(1-r u)},\quad B(u)\equiv 0,\quad C(u)=-%
\frac{1}{4}u^2.
\end{equation*}
Hence the system \eqref{ode}, determining coefficients of function $f_k$ in %
\eqref{fg}, takes the following form: 
\begin{align}  \label{odesw}
& G^{\prime \prime }_{k,k}- \frac{k(k-1)(1-r^2)}{(1-u^2)(1-r u)^2}
G_{k,k}=0,\qquad G^{\prime \prime }_{k,k-1}- \frac{(k-1)(k-2)(1-r^2)}{%
(1-u^2)(1-r u)^2}G_{k,k-1}=0, \\[3mm]
& G^{\prime \prime }_{k,k-s}-\frac{(k-s)(k-s-1)(1-r^2)}{(1-u^2)(1-r u)^2}%
G_{k,k-s}=\frac{(k-s+1)(k-s+2)(1-ru)}{1-u^2} G_{k,k-s+2},  \notag
\end{align}
with $s=2,\ldots, k.$ In this case the general solution to the first
equation of the system \eqref{odesw} can be written in closed form: 
\begin{align}  \label{sol}
&G_{k,k}=c_1 G^1_{k,k}+c_2 G^2_{k,k},\quad\text{with}  \notag \\[3mm]
&G^1_{k,k}=\frac{(1-u^2)}{(1-ru)^{k-1}}\bigl(u^{k-2}+R_{k,k-3}u^{k-3}+%
\ldots+R_{k,1} u+R_{k,0}\bigr), \\[3mm]
&G^2_{k,k}=\frac{Q_{k,k-1}u^{k-1}+Q_{k,k-1}u^{k-1}+\ldots+Q_{k,1} u+Q_{k,0}}{%
(1-ru)^{k-1}}+\Psi_k G^1_{k,k}\ln{\left(\frac{1-u}{1+u}\right)},  \notag
\end{align}
where all coefficients $R_{k,i}$, $Q_{k,i}$ and $\Psi_k$ can be determined
uniquely under substitution of \eqref{sol} in \eqref{odesw}. One can
similarly obtain general solution to the second equation in \eqref{odesw}.
Then all right hand sides in the rest of equations are completely determined
and one is able to construct general solution of the system \eqref{odesw}.

Looking at expressions in \eqref{sol} one can conclude that the complexity
of formulae increases with growth of $k$. For example in the case $k=3$ we
have 
\begin{equation}  \label{f3}
f_3(u,v)=G_{3,3}(u)v^3+G_{3,2}(u)v^2+G_{3,1}(u)v+G_{3,0}(u),
\end{equation}
where 
\begin{align*}
G_{3,3}&=\frac{c_{31}}{2}\frac{\Gamma^1_{3,3}}{(1-ru)^2}+\frac{c_{32}}{2}%
\left\{\frac{\Gamma^2_{3,3}}{(1-ru)^2}-\frac{3 \Gamma^3_{3,3}}{(1-ru)^2}\ln{%
\left(\frac{1-u}{1+u}\right)}\right\}, \\[3mm]
G_{3,2}&=c_{21}\left\{\frac{(1-u^2)}{(1-ru)}\right\}+\frac{c_{22}}{2}\left\{%
\frac{\Gamma^1_{3,2}}{(1-ru)}+\frac{\Gamma^2_{3,2}}{(1-ru)}\ln{\left(\frac{%
1-u}{1+u}\right)}\right\}, \\[3mm]
G_{3,0}&=c_{01}+c_{02}u+c_{21}u^2+c_{22}\left\{\frac{u}{2}%
(1+3r^2)+\Gamma^1_{3,0}\ln{(1-u)}+\Gamma^2_{3,0}\ln{(1+u)}\right\}
\end{align*}
and 
\begin{align*}
G_{3,1}&=c_{11}+c_{12}u+\frac{c_{31}}{r^3}\biggl\{\Gamma^1_{3,1}\ln(1-ru)+%
\Gamma^2_{3,1}\biggr\} \\
&-\frac{3}{4}\frac{c_{32}}{ r^3}\Gamma^3_{3,1}\biggl\{\mathrm{Li}_2\left(%
\frac{r(1-u)}{r-1}\right)-\mathrm{Li}_2\left(\frac{r(1+u)}{r+1}\right)%
\biggr\} \\[1mm]
&-\frac{3}{4}\frac{c_{32}}{ r^3}\Gamma^4_{3,1}\biggl\{\ln{(1+u)}\ln{\left(%
\frac{1-ru}{1+r}\right)}-\ln{(1-u)}\ln{\left(\frac{1-ru}{1-r}\right)}\biggr\}
\\[1mm]
&-\frac{3}{4}\frac{c_{32}}{ r^3}\biggl\{\Gamma^5_{3,1}+\Gamma^6_{3,1}\ln{%
(1-u)}+\Gamma^7_{3,1}\ln{(1+u)}+\Gamma^8_{3,1}\ln{(1-ru)}\biggr\}.
\end{align*}
Here $\mathrm{Li}_2(z)$ -- dilogarithm or Spence's function, defined as 
\begin{equation*}
\mathrm{Li}_2(z)=-\int\limits_0^z\frac{\ln{(1-t)}}{t}dt
\end{equation*}
for complex variable $z$; functions $\Gamma^i_{j,k}$ are polynomials in $u$
with coefficients being polynomials in $r$. All $\Gamma^i_{j,k}$ can be
found in explicit form by substitution of $G_{3,i}$ into the system %
\eqref{odesw} and listed in Appendix A.

Thus, one obtains solution in implicit form by formulae \eqref{t} and %
\eqref{x} using expression \eqref{f3}: 
\begin{align}  \label{t3}
t&=4 c_{21}+c_{22}\left\{\frac{T_{3,1}}{1-u^2}+T_{3,2}\ln{(1-u)}+T_{3,3}\ln{%
(1+u)}\right\} \\[1mm]
&+c_{31}\left\{\frac{12(u-r)v}{1-ru}\right\}+c_{32}\left\{\frac{T_{3,4}}{%
(1-ru)(1-u^2)}+\frac{T_{3,5}}{1-ru}\ln{\left(\frac{1-u}{1+u}\right)}%
\right\},\qquad\qquad\quad  \notag
\end{align}%
\\[3mm]
\begin{align}  \label{x3}
x&=2c_{22}\left\{\frac{v(1-ru)}{1-u^2}\right\}+c_{31}\left\{\frac{X_{3,1}}{%
r(1-ru)^2}+\frac{X_{3,2}}{r^2}\ln{(1-ru)}\right\}  \notag \\[1mm]
&+c_{12}+\frac{3}{2}\frac{c_{32}}{r^2}X_{3,3}\biggl\{\ln{(1+u)}\ln{\left(%
\frac{1-ru}{1+r}\right)}-\ln{(1-u)}\ln{\left(\frac{1-ru}{1-r}\right)}\biggr\}
\\[1mm]
&+\frac{3}{2}\frac{c_{32}}{r^2}\biggl\{X_{3,4}\ln{(1-ru)}+\frac{X_{3,5}}{%
(1-ru)^2}\ln{\left(\frac{1-u}{1+u}\right)}+\frac{X_{3,6}}{(u^2-1)(1-ru)^2}%
\biggr\}  \notag \\[1mm]
&+\frac{3}{2}\frac{c_{32}}{r^2}X_{3,7}\biggl\{\mathrm{Li}_2\left(\frac{r(1-u)%
}{r-1}\right)-\mathrm{Li}_2\left(\frac{r(1+u)}{r+1}\right)\biggr\}+\frac{3}{2%
}\frac{c_{32}}{r^2}X_{3,8}\ln{\Bigl((1+u)(1-u)\Bigr)} ,  \notag
\end{align}%
\\[1mm]
where $T_{i,j}$ and $X_{i,j}$ are polynomials in $u$ and $v$ with
coefficients being polynomials in $r$ (see Appendix A). Here it should be
also mentioned, that terms with coefficients $c_{21}$ and $c_{22}$ form the
solution for $k=2$. Therefore, the solution for every $k$ contains already
found solutions for lower values of $k$.

In the theory of stratified fluid dynamics the Boussinesq approximation is
relevant from physical point of view \cite{gavsukhmak}. Then for $r\ll 1$
the leading order term of \eqref{t3}, \eqref{x3} has the form: 
\begin{align*}
t&=4 c_{21}+c_{22}\left\{\frac{2 u}{1-u^2}+2\ln{\left(\frac{1-u}{1+u}\right)}%
\right\}+12c_{31}u v+6 c_{32}v\left\{\frac{2-3u^2}{u^2-1}+3u\ln{\left(\frac{%
1-u}{1+u}\right)}\right\} \\[3mm]
x&=\,\,\,c_{12}+c_{22}\left\{\frac{2v}{u^2-1}\right\}-3c_{31}\biggl\{%
v^2(1+u^2)+u^2\biggr\} \\[1mm]
&\qquad\quad\,+ \frac{3}{2}c_{32}\left\{\frac{-3u+uv^2+3u^3(1+v^2)}{u^2-1}-%
\frac{1-4u^2-3v^2+3u^4(1+v^2)}{u^2-1}\ln{\left(\frac{1-u}{1+u}\right)}%
\right\}.
\end{align*}%
\\[1mm]

\section{Liquid Layer on the Rotating Cylinder Surface}

Now consider the system 
\begin{equation}  \label{zh}
S_t=-\Bigl\{\Pi\ln{S}\Bigr\}_x,\quad \Pi_t=-\left\{\frac{\Pi^2}{2S}-\frac{1}{%
2}\Omega^2S-\Omega\Pi\ln{S}\right\}_x,\quad S=R^2,
\end{equation}
where $R$ is the free surface equation; $\Pi$ is the velocity circulation
and $\Omega$ is vorticity constant. The system \eqref{zh} arise in
describing of a thin liquid layer evolution on the rotating infinite
cylinder \cite{MZH}. This case corresponds to Hamiltonian density 
\begin{equation}  \label{hzh}
h(u,v)=-\frac{1}{2}v^2\ln{u}+\frac{1}{2}v\,\Omega\, u\ln{u}-\frac{1}{8}%
\Omega^2 u^2\ln{u}+\frac{1}{4}\Omega^2 u^2
\end{equation}
with $u=S$ and $v=\Pi+\Omega S/2$. As for the two-layer shallow water system
there are two limit cases \cite{MZH}:

\begin{itemize}
\item \textbf{the first case}, $\Omega\to 0$, corresponds to non-rotating
cylinder. For reduced system 
\begin{equation*}
S_t=-\Bigl\{\Pi\ln{S}\Bigr\}_x,\quad \Pi_t=-\left\{\frac{\Pi^2}{2S}\right\}_x
\end{equation*}
one is able to calculate Riemann invariants in explicit form: 
\begin{equation*}
r_1=\frac{\ln{\Pi}+\sqrt{-2\ln{S}}}{2},\quad r_2=\frac{\ln{\Pi}-\sqrt{-2\ln{S%
}}}{2}.
\end{equation*}

\item \textbf{the second case}, when parameter $\Omega\gg 1$, is related to
the so-called ``overturned'' shallow water system. Suppose free surface
variable $S$ and parameter $\Omega$ admit the following asymptotic
representation: 
\begin{equation*}
S=1+\varepsilon \Phi+O(\varepsilon^2),\quad \Omega=\sqrt{2/\varepsilon},
\end{equation*}
where $\varepsilon\ll 1$. The asymptotics corresponds to the situation, when
cylinder rotates very fast ($\Omega\gg 1$ for small $\varepsilon$) and free
surface disturbances are small as well. Therefore at the leading order in $%
\varepsilon$ one obtains the system 
\begin{equation*}
\Phi_t=-(\Phi\Pi)_x,\quad \Pi_t=-(\Pi^2/2-\Phi)_x,
\end{equation*}
which coincides with shallow water model up to gravity force direction.
\end{itemize}

For the Hamiltonian density \eqref{hzh} explicit expressions for $A$, $B$
and $C$ are as follows: 
\begin{equation*}
A(u)=-\frac{1}{2}\ln{u},\quad B(u)=\frac{1}{2}\Omega\,u\ln{u},\quad C(u)=-%
\frac{1}{8}\Omega^2 u^2\ln{u}+\frac{1}{4}\Omega^2 u^2
\end{equation*}
and the system \eqref{ode} is of the form 
\begin{align}  \label{odezh}
& G^{\prime \prime }_{k,k}+ \frac{k(k-1)}{2u^2 \ln{u}} G_{k,k}=0,\qquad
G^{\prime \prime }_{k,k-1}+ \frac{(k-1)(k-2)}{2u^2 \ln{u}}G_{k,k-1}=-\frac{%
\Omega k(k-1)}{2u\ln{u}}G_{k,k},  \notag \\[3mm]
& G^{\prime \prime }_{k,k-s}+\frac{(k-s)(k-s-1)}{2u^2 \ln{u}}G_{k,k-s}=-%
\frac{\Omega (k-s)(k-s+1)}{2u\ln{u}}G_{k,k-s+1} \\[1mm]
&\qquad\qquad\qquad\qquad\quad\quad\,\,\,\,+\frac{(k-s+1)(k-s+2)}{2\ln{u}}%
\left\{-\frac{1}{4}\Omega^2+\frac{1}{2}\Omega^2\ln{u}\right\} G_{k,k-s+2}, 
\notag
\end{align}
with $s=2,\ldots, k.$ After introducing a new variable $w=\ln{u}$ and making
substitution $G_{k,k}(u)=H_{k,k}(w)$ the system above transforms to the
following one 
\begin{align}  \label{odezh1}
& H^{\prime \prime }_{k,k}-H^{\prime }_{k,k}+\frac{k(k-1)}{2w} H_{k,k}=0, 
\notag \\[3mm]
&H^{\prime \prime }_{k,k-1}-H^{\prime }_{k,k-1}+ \frac{(k-1)(k-2)}{2w}%
H_{k,k-1}=-\frac{\Omega k(k-1)}{2w}e^{w}H_{k,k}, \\[3mm]
& H^{\prime \prime }_{k,k-s}-H^{\prime }_{k,k-s}+\frac{(k-s)(k-s-1)}{2w}%
H_{k,k-s}=-\frac{\Omega (k-s)(k-s+1)}{2w}e^w H_{k,k-s+1}  \notag \\[1mm]
&\qquad\qquad\qquad\qquad\qquad\qquad\,\,\,\,-\frac{1}{4}%
\Omega^2(k-s+1)(k-s+2)\frac{(1-2w)e^{2w}}{2w}H_{k,k-s+2},  \notag
\end{align}
for $s=2,\ldots, k.$ The first equation in \eqref{odezh1} is a particular
case of the following ODE 
\begin{equation*}
w H^{\prime \prime }+(\alpha+1-w)H^{\prime }+n H=0,
\end{equation*}
whose polynomial solutions are called generalized Laguerre polynomials 
\begin{equation*}
L^{(\alpha)}_n(w)=\frac{w^{-\alpha}}{n!}e^{w}\frac{d^n}{dw^n}%
\left(e^{-w}w^{n+\alpha}\right),
\end{equation*}
where $n$ is non-negative integer and $\alpha$ is arbitrary real number (see 
\cite{liz}). Thus the general solution to the first equation in %
\eqref{odezh1} has the form 
\begin{equation*}
H_{k,k}(w)=c_1 L^{(-1)}_{k(k-1)/2}(w)+c_2 H^{(2)}_{k,k}(w),
\end{equation*}
where function $H^{(2)}_{k,k}$ is defined by Liouville's formula. Thus, all
right hand sides of the system \eqref{odezh1} are determined and one is able
to get it's general solution. In the simpliest case $k=3$ for coefficients $%
H_{3,i}$ we have 
\begin{align*}
H_{3,3}&=c_{31}P^1_{3,3}+c_{32}\biggl\{e^w P^2_{3,3}+\mathrm{Ei}(w)P^1_{3,3}%
\biggr\}, & &H_{3,2}=c_{21}w+c_{22}\biggl\{w\mathrm{Ei}(w)-e^w\biggr\}%
+c_{31}\Omega e^{w}P^1_{3,2} \\
&\, & \,&+c_{32}\Omega\biggl\{e^{2w}P^2_{3,2}+e^{w}\mathrm{Ei}(w)P^3_{3,2}+2w%
\mathrm{Ei}(2w)\biggr\},
\end{align*}
\begin{align*}
H_{3,1}&=c_{11}e^w+c_{12}+c_{21}e^w\Omega(1-w)+c_{22}\Omega\biggl\{%
e^{2w}+e^w(2-w)\mathrm{Ei}(w)-2\mathrm{Ei}(2w)\biggr\} \\[2mm]
&\,+c_{31}e^{2w}\Omega^2 P^1_{3,1}+c_{32}\Omega^2\biggl\{%
e^{3w}P^2_{3,1}+e^{2w}P^3_{3,1}\mathrm{Ei}(w)+2e^{w}(2-w)\mathrm{Ei}(2w)-%
\frac{135}{64}\mathrm{Ei}(3w)\biggr\},
\end{align*}
\begin{align*}
H_{3,0}&=c_{01}e^w+c_{02}+c_{21}e^{2w}\Omega^2P^1_{3,0}+c_{31}e^{3w}\Omega^3
P^3_{3,0} \\[2mm]
&\qquad+c_{22}\Omega^2\biggl\{-\frac{1}{4}e^{3w}+e^{2w}P^2_{3,0}\mathrm{Ei}%
(w)+e^w\mathrm{Ei}(2w)-\frac{1}{2}\mathrm{Ei}(3w)\biggr\} \\[2mm]
&\qquad+c_{32}\Omega^3\biggl\{e^{4w}P^4_{3,0}+e^{3w}P^5_{3,0}\mathrm{Ei}%
(w)+e^{2w}P^6_{3,0}\mathrm{Ei}(2w)+\frac{135}{128}e^w\mathrm{Ei}(3w)-\frac{8%
}{27}\mathrm{Ei}(4w)\biggr\},
\end{align*}
where $P^i_{j,k}$ are polynomials with respect to $w$ listed in Appendix B
and $\mathrm{Ei}(z)$ is exponential integral function defined for complex
variable $z$ as follows 
\begin{equation*}
\mathrm{Ei}(z)=\int\limits^{z}_{-\infty}\frac{e^t\,dt}{t}.
\end{equation*}
Thus, formulae \eqref{t} and \eqref{x} give the solution in implicit form: 
\begin{align}  \label{tzh}
t&=2c_{21}+2c_{22}\biggl\{\mathrm{li}(u)-\frac{u}{\ln{u}}\biggr\}+c_{32}\,v%
\biggl\{20 u-4u\ln{u}-\frac{8u}{\ln{u}}+\mathrm{li}(u)T_{3,4}\biggr\} \\%
[1.5mm]
&\quad+c_{31}\biggl\{v\,T_{3,1}-\Omega\,u\, T_{3,2}\biggr\}+\Omega\,c_{32}%
\biggl\{32 \,\mathrm{Ei}(2 \ln{u})-9 u^2+u^2 \ln{u}-\frac{2 u^2}{\ln{u}}- u\,%
\mathrm{li}(u)T_{3,3}\biggr\},  \notag
\end{align}

\begin{align}
x& =\Omega ^{2}c_{32}\biggl\{-\frac{25}{16}u^{2}+4\mathrm{Ei}(2\ln {u})-%
\frac{u^{2}}{8\ln {u}}+\frac{1}{2}u^{2}\ln {u}-\frac{1}{16}u^{2}\ln ^{2}{u}%
+u\,\mathrm{li}(u)X_{3,6}\biggr\}  \notag  \label{xzh} \\[1.5mm]
& +c_{32}\biggl\{v^{2}\left( -\frac{1}{4}+\frac{1}{\ln {u}}-\frac{1}{4}\ln {u%
}+\frac{\mathrm{li}(u)}{u}X_{3,3}\right) +\Omega v\left( \frac{3}{2}u+\frac{u%
}{\ln {u}}X_{3,4}+\mathrm{li}(u)X_{3,5}\right) \biggr\} \\[1.5mm]
& +c_{21}\Omega +c_{22}\biggl\{2\Omega \,\mathrm{li}(u)-\frac{2v-\Omega u}{%
\ln {u}}\biggr\}+c_{31}\biggl\{\frac{3v^{2}\left( \ln ^{2}{u}-6\right) }{u}+%
\frac{3}{2}v\,\Omega \,X_{3,1}+\frac{3}{4}u\,\Omega ^{2}\,X_{3,2}\biggr\}, 
\notag
\end{align}%
where functions $T_{i,j}$, $X_{i,j}$ are polynomials with respect to $\ln {u}
$ (see Appendix B) and $\mathrm{li}(u)=\mathrm{Ei}(\ln {u})$ is logarithmic
integral function.

\section{Conclusion}

It is well known that two-component two-dimensional hydrodynamic type
systems are integrable by the hodograph method. Usually, this means, that
corresponding system of two linear equations with variable coefficients can
be solved explicitly, i.e. one can construct a general solution
(parameterised by two arbitrary functions of a single variable), or at least
one can derive two infinite series of particular solutions. One such an
example is a polytropic gas (where a pressure depends on a density only).
The second example was found by P. Olver and Ya. Nutku \cite{on}. Here we
presented a third example, which has applications in dynamic of fluid. This
case contains three arbitrary functions instead of a sole function in two
above investigated cases. This advantage allows to apply our results to a
wide range of hydrodynamic type systems.

\section{Acknowledgements}

ZVM and MVP are grateful to A.A. Chesnokov, A.K. Khe, N.I. Makarenko, 
A.M. Kamchatnov and M.Yu. Zhukov for very important comments, remarks and helpful
conversations. ZVM was supported by the project No 2.3.1.2.12 (code FWGG-2021-0011). MVP was supported by the project No 0023-2019-0011.

\newpage

\section{Appendix A}

In this section we list explicit formulae $\Gamma^i_{j,k}$ for $G_{3,i}$ in %
\eqref{f3} and $X_{i,j}$, $T_{i,j}$ in \eqref{x3}, \eqref{t3}: 
\begin{align*}
\Gamma^1_{3,3}&=(r-u)(u^2-1),\quad \Gamma^2_{3,3}=u^2(2 r^4-3 r^2+3)-u(3r
+r^3)-2+6 r^2-2 r^4, \\
\Gamma^3_{3,3}&=(r^2-1)\Gamma^1_{3,3}. \\
\,&\, \\
\Gamma^1_{3,2}&=(1+r^2)u-2r,\quad \Gamma^2_{3,2}=(r^2-1)(u^2-1), \\
\,&\, \\
\Gamma^1_{3,1}&=6(r^2-1)(ru-1),\quad \Gamma^2_{3,1}=-3r^2u(u+2r)+6ru,\quad
\Gamma^3_{3,1}=\Gamma^4_{3,1}=(r^2-1)\Gamma^1_{3,1}, \\
\Gamma^5_{3,1}&=2r^2u(3r^2+4r^3-3)+6(r^2-1)^2\ln{\left(\frac{1+r}{1-r}\right)%
}, \\
\Gamma^6_{3,1}&=r(r-1)(u-1)(8r^3+r^2-9r-6+3ru+3r^2u), \\
\Gamma^7_{3,1}&=r(r+1)(u+1)(8r^3-r^2-9r+6-3ru+3r^2u), \\
\Gamma^8_{3,1}&=-4r(ru-1)(3-5r^2+2r^4), \\
\,&\, \\
T_{3,1}&=2u(1+r^2)-4r,\quad T_{3,2}=(r^2-1),\quad T_{3,3}=r^2+1,\quad
T_{3,4}=6v\,\Gamma^2_{3,3}, \\
T_{3,5}&=-9v(1-r^2)(u-r), \\
\,&\, \\
X_{3,1}&=3r v^2 (u^2-2ru+1),\quad X_{3,2}=-6(1-r^2),\quad
X_{3,3}=-3(r^2-1)^2, \\
X_{3,4}&=2r (3-5r^2+2r^4), \\
X_{3,5}&=u^3r(6r^4-6r^2)+u^2r(12r-10r^3+2r^5+(3r-3r^3)v^2) \\
&+ur(-6+2r^2-4r^4+(6r^4-6r^2)v^2)+rv^2(3r-3r^3)+r(2r+2r^3), \\
X_{3,6}&=u^4(6r^3-6r^5)+u^3(-12r^2+12r^4+(-3r^2+3r^4-4r^6)v^2) \\
&+u^2(6r-12r^3+6r^5+(6r^3+6r^5)v^2)+u(12r^2-12r^4+(-r^2-15r^4+4r^6)v^2) \\
&+v^2(6r^3-2r^5)+6r^3-6r, \\
X_{3,7}&=3(r^2-1),\quad X_{3,8}=-3r+5r^3.
\end{align*}

\section{Appendix B}

In this section we list explicit formulae $P^i_{j,k}$ for $H_{3,i}$ and $%
X_{i,j}$, $T_{i,j}$ in \eqref{xzh}, \eqref{tzh}: 
\begin{align*}
P^1_{3,3}&=6w-6w^2+w^3,\quad P^2_{3,3}=-\frac{1}{6}+\frac{5}{12}w-\frac{1}{12%
}w^2, &  \\
\,&\, &  \\
P^1_{3,2}&=-\frac{33}{2}w+\frac{15}{2}w^2-\frac{3}{4}w^3,\quad P^2_{3,2}=-%
\frac{1}{8}-\frac{9}{16}w+\frac{1}{16}w^2,\quad P^3_{3,2}=-\frac{11}{8}w+%
\frac{5}{8}w^2-\frac{1}{16}w^3, &  \\
\,&\, &  \\
P^1_{3,1}&=\frac{3}{16}(-121+130w-42w^2+4w^3),\quad P^2_{3,1}=-\frac{1}{32}%
(16-19w+2w^2),\quad P^3_{3,1}=\frac{1}{12}P^1_{3,1}, \, & \, \\
P^1_{3,0}&=\frac{1}{4}(w-2),\quad P^2_{3,0}=P^1_{3,0},\quad P^3_{3,0}=-\frac{%
1}{288}(-835+858w-234w^2+18w^3), &  \\
P^4_{3,0}&=\frac{1}{576}(-34-36w+3w^2),\quad P^5_{3,0}=\frac{1}{12}%
P^3_{3,0},\quad P^6_{3,0}=2P^1_{3,0}, &  \\
\,&\, &  \\
T_{3,1}&=36-36\ln{u}+6\ln^2{u},\quad T_{3,2}=33-15\ln{u}+\frac{3}{2}\ln^2{u}%
,\quad T_{3,3}=22 -10\ln{u}+\ln^2{u}, &  \\
T_{3,4}&=24-24\ln{u}+4\ln^2{u}, &  \\
\,&\, &  \\
X_{3,1}&=\ln^3{u}-2\ln^2{u}-12\ln{u}+12,\quad X_{3,2}=-50 +\ln^3{u}-9\ln^2{u}%
+32\ln{u}, &  \\
X_{3,3}&=-\frac{3}{2}+\frac{1}{4}\ln^2{u},\quad X_{3,4}=-\frac{1}{4}+\frac{1%
}{8}\ln^2{u}-\frac{1}{8}\ln^3{u}, &  \\
X_{3,5}&=\frac{3}{2}-\frac{3}{2}\ln{u}-\frac{1}{4}\ln^2{u}+\frac{1}{8}\ln^3{u%
},\quad X_{3,6}=-\frac{25}{8}+2\ln{u}-\frac{9}{16}\ln^2{u}+\frac{1}{16}\ln^3{%
u}. & 
\end{align*}

\end{document}